\documentstyle[11pt,paspconf,epsf]{article}

\begin{document}

\title{The Sagittarius Dwarf Galaxy Survey (SDGS): constraints on the Star
Formation History of the Sgr dSph}

\author{M. Bellazzini, F.R. Ferraro} 

\affil{Osservatorio Astronomico di Bologna, via Zamboni 33, 40126 Bologna,
ITALY}

\author{and R. Buonanno}
\affil{Osservatorio Astronomico di Roma, Monte Porzio Catone, Roma, ITALY}

\begin{abstract}
The main characteristics of a wide photometric survey of the Sgr dwarf 
spheroidal galaxy are shortly presented. 
V and I photometry has been obtained for $\sim 90000$ 
stars toward Sgr and for $\sim 9000$ stars in a region devoid of Sgr stars
(for decontamination purposes).

The full potential of this large database is far from being completely explored.
Here we present only preliminary results from the analysis of statistically
decontaminated Color Magnitude Diagrams, trying to set a scheme of the Star 
Formation History of the Sgr Galaxy. A scenario is proposed in which star 
formation in Sgr began very
early and lasted for several Gyr with progressive chemical enrichment of the
Inter Stellar Medium (ISM). Nearly 8 Gyr ago the star formation rate abruptly
decreased, perhaps in coincidence with the event that led to the gas depletion
of the galaxy. A very small rate of star formation continued until relatively
recent times ($\sim 1$ Gyr ago).

\end{abstract}




\section{Introduction}

The Sagittarius Dwarf Spheroidal galaxy (Sgr dSph) is
the nearest Milky Way satellite. The detailed study of this stellar system can 
potentially provide important insights on topics ranging from evolution of 
dwarf galaxy to the formation of the Milky Way halo (see Ibata et al 1997 and 
Bellazzini, Ferraro \& Buonanno 1998, for details and references).

Despite of its relative nearness Sgr represents a noticeable challenge from an
observational point of view because of its very low surface brightness (one
needs to sample wide fields to adequately populate the brightest part of the
Color Magnitude Diagram) and of the strong contamination by foreground stars
from the Milky Way bulge and disk (statistical decontamination of Color
Magnitude Diagrams (CMD) is necessary).

In the attempt of overcoming both problems we planned a large photometric
survey of the galaxy. Three large ($9 \times 35 ~arcmin^2$) fields placed
in different regions of Sgr dSph have been
observed in V and I passband. Calibrated photometry has been obtained for 
nearly 90000 stars in these fields. Addictional $\sim 8000$ stars have been
observed in a control field devoid of Sgr stars, for statistical 
decontamination purposes.
Observations have been carried out at the ESO-NTT and data has been reduced
with the ROMAFOT package.

In table 1 the main characteristics of the observed fields are 
reported\footnote{The original Color Magnitude Diagrams are not 
presented in the written version of the contribution, because of the limited 
space available.}.
The best photometric quality has been achieved for the SGR34 and SGRWEST fields.
The limiting magnitude for these fields is around $V\sim 23$ while in the case
of the SGR12 and GAL fields it is some 1 mag brighter.
 The GAL field samples the foreground population at the same galactic
latitude of SGR12 but outside the body of the Sgr dSph. A deeper (V,I)
photometry of the GAL region has been obtained by MUSKKK and has been adopted as 
final control field sample for decontamination of the faintest part of our CMDs.

\begin{table}
\caption{Names and positions of SDGS Fields.} \label{tbl-1}
\begin{center}\scriptsize
\begin{tabular}{ccccc}
Name    & $l^o$ & $b^o$ & Field Size        &$N_{\star}$\\ 
\tableline
SGR34   & 6.5 & -16 & $35^{'} \times 9^{'}$ &22603\\
SGR12   & 6   & -14 & $35^{'} \times 9^{'}$ &25793\\
SGRWEST & 5   & -12 & $35^{'} \times 9^{'}$ &41462\\
GAL     & 354 & -14 & $21^{'} \times 9^{'}$ &8336 \\
\end{tabular}
\end{center}
\end{table}

\begin{figure}  
\plotone{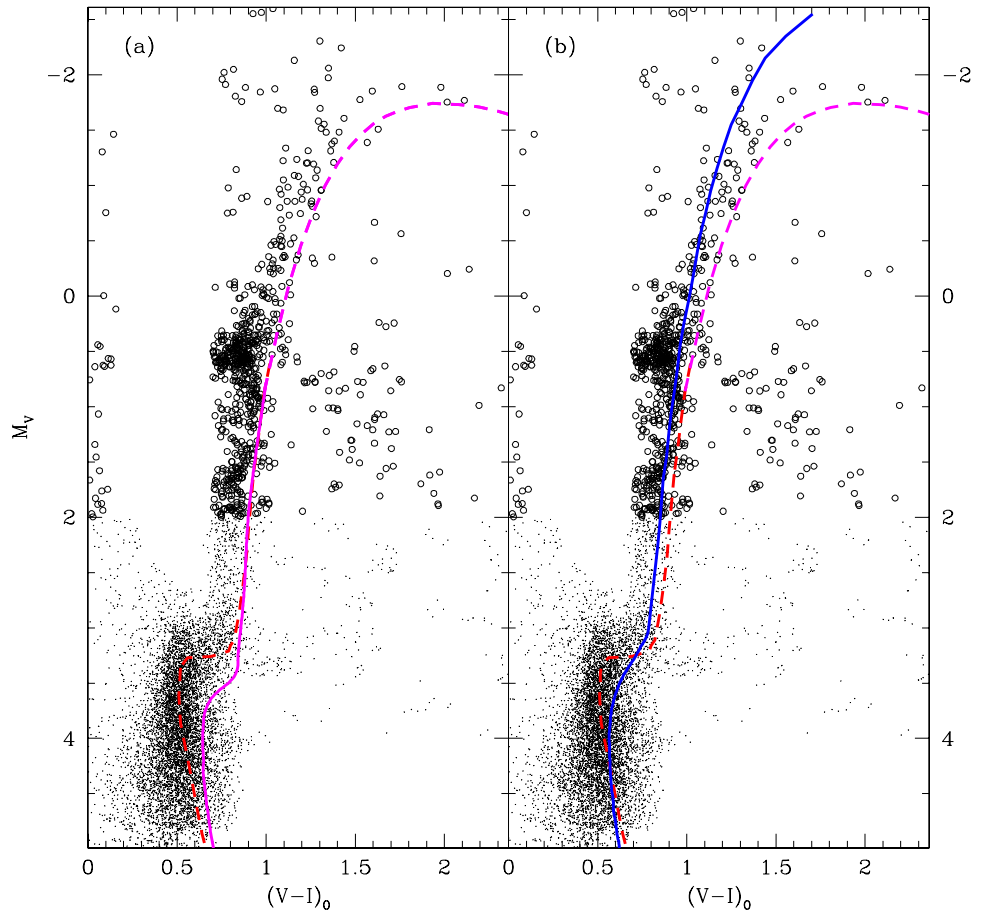}
\caption{The decontaminated CMD of the SGR34 field is reported in both panels.
Two different kind of symbols have been adopted to represent stars brighter
than $M_V=2$ (open circles) and stars fainter than $M_V=2$ (tiny points), in
order to let the superposed ridge lines to remain visible despite of the large
number of stars present in the TO region. In panel (a) the ridge lines of 47
Tuc (continuous line; classical old globular with [Fe/H]=-0.7) and of Pal 1
(dashed line; the same metallicity than 47 Tuc but $\sim 4 ~Gyr$ younger) are
also reported). From this panel it can be concluded that the more metal rich 
stars in the SGR34 region are significanly younger than 47 Tuc and have an age
similar to Pal 1, whose ridge line provides an excellent fit of the TO region.
The continuous ridge line reported in panel (b) is that of M 3 (old and with
[Fe/H]=-1.3), also consistent with the observed CMD of the TO region. It can be
noted that the two reported ridge lines [M 3 and Pal 1 (dashed line)] provide a
global fit to the observed CMD, simultaneously reproducing the TO region and
the spreaded RGB.}
\end{figure}

A deep analysis of the obtained CMDs has been performed before attempting
statistical decontamination and a first paper dealing with distance, reddening,
degree of foreground contamination in the various fields and many other topics
has been submitted to MNRAS (Bellazzini, Ferraro \& Buonanno 1998, hereafter
PAP-I). For the present purposes it is useful to recall two of the results of
PAP-I, i.e. (1) the reddening differences between the various SDGS fields are
tiny and (2) there are evidences for the presence of a population as metal poor
as $[Fe/H]=-2.0$ associated with the Sgr dSph.

Here we just present preliminary results from the analysis of the statistically
decontaminated CMDs of the SGR34 and SGR12 fields, providing some constraint on
the age and metallicity distribution of Sgr stars and proposing a general
scheme for the Star Formation History in this galaxy.

\section{Metallicity}

Comparing the RGB stars distribution (from decontaminated CMDs) with 
ridge lines from template globular clusters it is concluded that the stars 
belonging to the main population of the Sgr dSph span a wide metallicity range 
(more than 1 dex), from $[Fe/H]\sim-2.0$ to $[Fe/H]=-0.7$, the bulk of the
stars having $-1.3<[Fe/H]<-0.7$. Few stars more metal rich than $[Fe/H]=-0.7$
are probably present in the SGR12 field, confirming previous claims of a small
metallicity gradient toward the Sgr center of density (Sarajedini \& Layden
1995; Marconi et al 1997). 

\section{A possible SFH for the Sgr dSph} 

The decontaminated CMD of the SGR34 field is shown in fig. 1 [panel (a) and
(b)]. The obvious ``globular cluster -like'' features are typically associated 
with an intermediate-old main population (MUSKKK, Marconi et al 1997). 
Hereafter we will refer to this component as Sgr Pop A. 
The sparse Blue Plume of
stars around $M_V \sim 3$ and $(V-I)_0 \sim 0.7$ is usually interpreted as an
extended Main Sequence associated with more recent star formation
episodes\footnote{This interpretation is mainly based on the detection of
Carbon stars in Sgr, taken as a clear signature of the presence of young stars.
However, a significant fraction of the stars in the Blue Plume can well be
genuine Blue Straggler stars (see PAP-I).} (hereafter, Sgr Pop B).

\begin{figure}  
\plotone{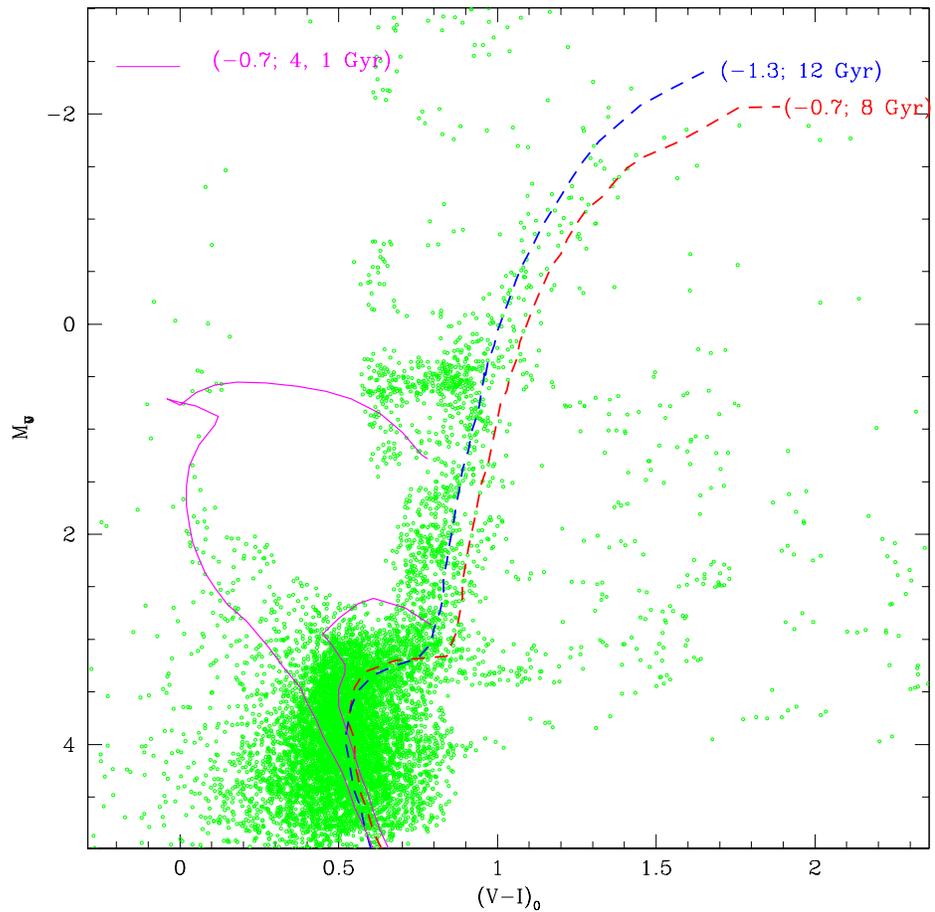}
\caption{Decontaminated CMD of the SGR34 field. Dashed lines are ``old''
isochrones, fitting the main population (Pop A). Age (in Gyr) and [Fe/H] values
are indicated in parenthesis. Continuous lines are (from left to right) a 1
Gyr isochrone and a 4 Gyr isochrone at [Fe/H]=-0.7, bracketing the Blue Plume
stars distribution. All the reported isochrones are from Bertelli et al (1994).
[note: The color versions of fig. 1 and fig. 2 are much clearer. Use
{\em gosthview} to look at the retrieved postscript files]}
\end{figure}

Till now, the standard way to deal with the age of Sgr Pop A was to assign an
average metallicity to the population and then deriving an age via isochrone
fitting of the TO region. 
The present day data are consistent with a single TO point (or, at least,
with few slightly different and unresolved TO points superposed in the same CMD
region). Stellar evolution teach us that populations of different metallicity
can share a common TO point only if they differ also in age, in the sense that
more metal rich populations have to be younger (i.e., the usual sense of any
well-behaved age-metallicity relation). We propose this kind of solution for
the SFH of Sgr Pop A as the most likely from many point of view (see
Bellazzini, Ferraro \& Buonanno 1999). This is the first attempt to provide a
global fit to both the TO region and the spreaded RGB of the Sgr CMD.

In panel (a) of fig. 1 the ridge lines of 47 Tuc (from Kaluzny et al 1998;
continuous line) and of Pal 1 (from Rosenberg et al 1998a; dashed line) are
superposed to the observed CMD. The two clusters have the same metallicity
([Fe/H]=-0.7; Rosenberg et al 1998b) but Pal 1 is younger than 47 Tuc 
by $\sim 4 Gyr$. 
While the ridge line of 47 Tuc clearly do not fit the TO region of Sgr an 
excellent fit is provided by Pal 1. At present, this is the best fit to the
whole Sgr CMD based on empirical (V,V-I) ridge lines. 
It is concluded that the more metal rich stars in SGR34 are
significantly younger than 47 Tuc and have nearly the same age of Pal 1.
However the ridge line of Pal 1 can't provide, obviously, a satisfying fit of
the spreaded RGB of the Sgr galaxy, a more metal poor component is also needed.

The continuous line reported in panel (b) is the ridge line of the M 3 globular
cluster (a classical old globular with [Fe/H]=-1.3; Ferraro et al. 1997), while
the dashed line is (again) the Pal 1 ridge line. The M 3 ridge line is not
unconsistent with the observed CMD in the TO region and provides a good fit for
a number of Sgr RGB stars. Coupling the two ridge lines  
a (qualitative) simultaneous fit of all the observed characteristics of the 
CMD of Pop A is provided\footnote{The ridge
line of Ter 8 (old and with [Fe/H]=-2.0; not shown here) is also consistent 
with the observed TO and fits the bluest RGB stars in the SGR34 Color Magnitude
Diagram}. The comparisons shown in fig. 1 strongly suggest that a spread in age
is coupled with the observed metallicity spread.   

The same scheme can be checked through comparison with theoretical isochrones,
as shown in fig. 2. 
The dashed lines superposed to the CMD of the SGR34 field are isochrones of 
([Fe/H]=-1.3; age=12 Gyr) and ([Fe/H]=-0.7; age=8 Gyr) respectively 
(see caption). 
Coupling the two, a
consistent solution can be found fitting both the TO region and the spreaded
RGB. Adding a ([Fe/H]=-2.0; age=16 Gyr) would complete the scheme, providing a
satisfying interpretation for all the properties of Pop A: a long lasting OLD
(age $>$ 8 Gyr) star formation event accompanied by progressive chemical
enrichment. This Star Formation History scheme is also supported by the Age
Metallicity Relation derived by Montegriffo et al (1998) for the globular
cluster system of the Sgr galaxy.

A clear-cut confirmation of the proposed scheme is not possible at the moment, 
given the accuracy of the SDGS data at the TO level. However HST photometry 
(Mighell at al. 1997) will certainly allow a much more accurate analysis of the 
TO region, checking the present hypothesis and providing more details to 
the scenario. 

The Blue Plume stars are nicely bracketed between a 1 Gyr isochrone and a 4 Gyr
isochrones, both at [Fe/H]=-0.7. The adoption of higher metallicity isochrones
would have the only effect of shifting the age range of Pop B to younger ages.
It is concluded that,
massive star formation in Sgr suddenly stopped several Gyrs ago (Pop A), while
more marginal episodes lasted until recent times (Pop B provides less than 
$10 \%$ of the whole Sgr dSph stellar content; PAP-I).

The proposed scheme can be applied to all the observed regions of the galaxy,
since the stellar content is found to be remarkably homogeneous over the whole
sampled scale ($\sim 3.5$ Kpc along the major axis).

\end{document}